\documentclass[lettersize,journal]{IEEEtran}
\usepackage{amsmath,amsfonts}
\IEEEoverridecommandlockouts
\pdfoutput=1
\usepackage{cite}
\usepackage{amsmath,amssymb,amsfonts}
\usepackage{algorithmic}
\usepackage{graphicx}
\usepackage{textcomp}
\usepackage{xcolor}

\usepackage{pifont}
\usepackage{amsmath,amssymb,amsfonts}
\usepackage{enumerate}
\usepackage{subfigure}
\usepackage{times}
\usepackage{float}
\usepackage{cite}
\usepackage{color}

\def\BibTeX{{\rm B\kern-.05em{\sc i\kern-.025em b}\kern-.08em
    T\kern-.1667em\lower.7ex\hbox{E}\kern-.125emX}}

\begin{document}

\title{Temporal-Assisted Dynamic Beampattern Optimization in Integrated Sensing and Communication Systems\\
}
\author{\IEEEauthorblockN{
                Shengcai~Zhou, \emph{Student~Member, IEEE}
		      Luping Xiang, \emph{Member, IEEE},
			and~Kun~Yang}, \emph{Fellow, IEEE}
       \thanks{The authors would like to thank the financial support of Quzhou Government (Grant No.: 2023D005), Natural Science Foundation of China (Grant No. 62132004 and Grant No. 62301122), MOST Major Research and Development Project (Grant No.: 2021YFB2900204), Sichuan Science and Technology Program (Grant No.: 2023NSFSC1375). \textit{(Corresponding author: Luping Xiang.)}}
        \thanks{Shengcai Zhou is with the Yangtze Delta Region Institute (Quzhou), University
of Electronic Science and Technology of China, Quzhou 324003, China, and
also the School of Information and Communication Engineering, University
of Electronic Science and Technology of China, Chengdu 611731, China,
email: 202222011112@std.uestc.edu.cn.}
        
        \thanks{Luping Xiang and Kun Yang are with the State Key Laboratory of Novel Software Technology, Nanjing University, Nanjing 210008, China, and School of Intelligent Software and Engineering, Nanjing University (Suzhou Campus), Suzhou, 215163, China, email:  luping.xiang@nju.edu.cn,  kunyang@nju.edu.cn.}
	}
\maketitle
\begin{abstract}
In this paper, an integrated sensing and communication (ISAC) system is investigated. Initially, we introduce a design criterion wherein sensing data acquired from the preceding time slot is employed for instantaneous optimal beamforming in the succeeding time slot, aiming to enhance the communication rate. Subsequently, the development of optimal beamforming is addressed, and a high-caliber suboptimal resolution is derived utilizing successive convex approximation (SCA) techniques combined with the iterative rank minimization (IRM) methodology. Our evaluations, grounded on numerical analyses, reveal that the communication rate of the introduced beamforming strategy surpasses that of conventional omnidirectional sensing and pilot based approaches.
\end{abstract}


\section{Introduction}
\IEEEPARstart{T}{he} integration of sensing and communication (ISAC) emerges as a focal approach \cite{10313997,9606831}, since the radio propagation is able to convey information generated from the transmitter and extract information based on scattered echoes \cite{https://doi.org/10.1049/smc2.12041}. The ISAC system allows a co-designed waveform to transmit information to the communication user while sensing the radar target. Sensing-assisted communication technology, which enables communication and sensing to cooperate to omit special pilot design , has attracted a lot of attention in recent years\cite{Fan2020RadarcommunicationSS}.

The study of \cite{8288677} proposed a shared deployment scheme that directly utilizes communication signals as dual-function waveforms, addressing the additional power loss issue caused by separate transmission of radar and communication signals  \cite{9916163}. In addition to the simple beamforming gain metrics \cite{8288677}, \cite{jing2022isac} also investigated the trade-offs between minimizing the Cramer-Rao Bound (CRB) and trajectory optimization. A further approach involves directly activating different numbers of transmitting antennas at the transmitter to achieve dynamic beampattern with varying beamwidths \cite{Trees2002OptimumAP}. Simultaneously, researchers in the Vehicle-to-Infrastructure (V2I) domain studied the sensing-assisted beam tracking rooted in radar echo data, bypassing the traditionally high-overhead pilot-based channel estimation \cite{9171304,9747255}. Initial V2I surveys used beam training to extract precise channel details, yet they still depended on pilot signals and failed to integrate radar-communication combined signals into a cohesive system. Furthermore, \cite{10098686} describes some popular framework protocol designs for UAVs, discussing performance variations based on factors such as quality of service requirements, target localization, and maneuverability.

This study delves into an optimal beamforming protocol supported by the ISAC framework. Contrary to prior research, our proposed A2G model employs an integrated waveform for persistent environmental sensing, subsequently employing this data to enhance communication. Additionally, we examine a optimization problem about the ramifications of imprecise sensing data on communication rates, manifested in beamforming through beam coverage breadth. An overview of this study's merits is encapsulated in Table \ref{contributions}. Key contributions from this research include:

\begin{table}[t]
\centering
\caption{Contrasting Our Contributions To The State-Of-Art}
\begin{tabular}{l|c|c|c|c|c }
 \hline 
 Contributions & \textbf{this work}  & \cite{9916163} & \cite{8288677} & \cite{9171304,9747255,10404096} & \cite{LIU2017331} \\ \hline\hline
 Continuous wave & \ding{52}  & \ding{51} & \ding{51} & \ding{51} &  \\
 \hline
 Integrated signal & \ding{52}  &  & \ding{51} & \ding{51} & \ding{51} \\
 \hline
 Beam coverage & \ding{52}  &  & \ding{51} &  &  \\
 \hline
 Sensing-assisted & \ding{52}  &  &  & \ding{51} & \ding{51} \\
 \hline
\end{tabular}

\label{contributions}
\end{table}

\begin{itemize}
\item A sensing-assisted communication protocol tailored for the ISAC system is introduced. This protocol designs the forthcoming transmit beampattern employing the channel data acquired from the preceding time slot to enhance the communication rate.

\item We employ the successive convex approximation (SCA) method \cite{Dinh2010LocalCO} to reframe the presented non-convex optimization challenge as a convex optimization problem. Subsequently, the iterative rank minimization (IRM) algorithm \cite{Sun2017TwoAF} is applied to derive high-caliber solutions.

\item Our simulation outcomes highlight the superiority of the devised protocol. When contrasted against the traditional omnidirectional sensing-based method and the pilot-based approach, there's an observable enhancement in throughput by 5.7$\%$ and 8.7$\%$, respectively.
\end{itemize}

The rest of this paper is organised as follows. Section II introduces the system model, Section III formulates the beam optimization problem. Section IV provides numerical results, and Section V summarizes this paper.
\section{System Model}

\begin{figure}[t] 
\centerline{\includegraphics[width=0.42\textwidth]{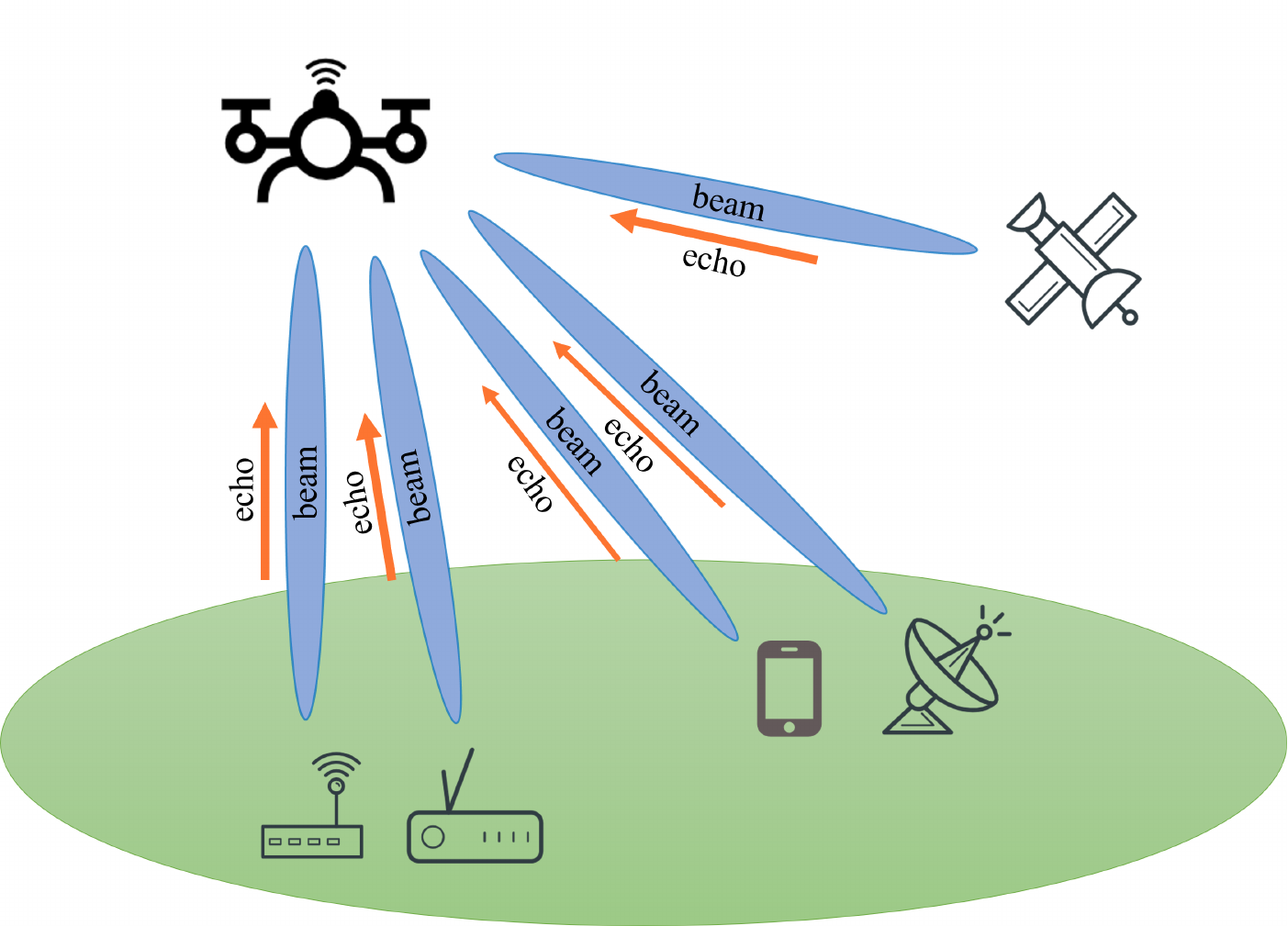}}
\caption{System model.}
\label{fig.scenario}

\end{figure}

As illustrated in Fig. \ref{fig.scenario}, an ISAC system is considered, comprising an aerial base station (ABS) equipped with a transmit uniform linear array (ULA) encompassing $N_t$ antennas and a distinct receive ULA consisting of $N_r$ antennas. This ABS is adept at sensing proximate single-antenna IoT devices and facilitating downlink communications to them. Without loss of generality, it is postulated that the ABS senses and communicates with the IoT devices exclusively over the line of sight (LoS) channels.

For the described system, it is assumed that the protocol endures for $N$ = 10 time intervals, each interval length is $ \Delta T$. Initially, ABS has imprecise information about $K$ devices in its vicinity, alongside an unknown number of devices without channel information, so it configures a multi-beam ISAC signal $\mathbf{s}(t)$, having $K$ dimensions, expressed as
\begin{equation}
\mathbf{s}_n(t)=[s_{1,n}(t),...,s_{K,n}(t)]^T,
\end{equation}
where the $k$-th signal $s_{k,n}(t)$ includes the communication information of the $k$-th device at time slot $n$, with independent circularly symmetric complex Gaussian (CSCG) random variables with zero mean and unit variance, represented by $s_{k,n}(t)\sim\mathcal{CN}(0,1)$.

Subsequent to the beamforming process, the resultant transmitted signal is
\begin{equation}
\mathbf{x}_n(t)=\mathbf{W}_n\mathbf{s}_n(t),
\end{equation}
where $\mathbf{W}_n\in \mathbb{C}^{N_t\times K}$ denotes the beamforming matrix (alternatively termed as the linear precoder).
\subsection{Omnidirectional Phase}
Initially, the ABS lacks knowledge regarding the channel information of the devices. Thus, the proposed protocol utilizes an omnidirectional beam for both sensing and communication with these devices. In this phase, we assign $n=1$.
\subsubsection{Communication Process}
According to literature \cite{9916163}, the channel vector from the ABS to the $k$-th device is represented as
\begin{equation}
\mathbf{h}_{k,n}=\frac{\alpha_0}{d_{k}}\mathbf{a}(\theta_k),
\end{equation}
where the term $\alpha_0$ signifies the path loss at a reference distance $d_0=1$ m. The distance between the ABS and the $k$-th device is $d_k$. The steering vector aimed at device $k$ is expressed as
\begin{equation}
\mathbf{a}(\theta_k)=[1,e^{j2{\pi}\frac{d}{\lambda}\sin{\theta_k}},...,e^{j2{\pi}\frac{d}{\lambda}(N_t-1)\sin{\theta_k}} ]^T,
\end{equation}
where $d$ represents the gap between two neighboring antennas and $\theta_k\in(-\pi,\pi)$ is the device's angular position.

Consequently, the signal received by the $k$-th device is
\begin{equation}
y_{k,n}(t)=\mathbf{h}_{k,n}^H\mathbf{x}_n(t)+z_C(t),
\end{equation}
with $z_C(t)\sim\mathcal{CN}(0,\sigma_C^2)$ symbolizing the additive white Gaussian noise (AWGN) at the receiver.

Given a unit total transmit power of $P_T$ and based on the principles of omnidirectional beampattern \cite{4516997}, the orthogonal nature of the signal $\mathbf{x}_n(t)$ is ensured. For scenarios where $K\geq N_t$, the spatial covariance matrix of the omnidirectional beam is
\begin{equation}
\mathbf{C}=\mathbb{E}[\mathbf{x}_n(t)\mathbf{x}^H_n(t)]=\mathbf{W}_n\mathbf{W}^H_n=\frac{P_T}{N_t}\mathbf{I}_{N_t},
\end{equation}
where $\mathbf{I}_{N_t}$ denotes the $N_t\times N_t$ identity matrix, the received SINR of the $k$-th device is represented as
\begin{equation}\label{omnisinr}
\gamma_{k,n}=\frac{\frac{1}{K}\mathbf{h}_k^H \mathbf{C} \mathbf{h}_k}{\frac{K-1}{K}\mathbf{h}_k^H \mathbf{C} \mathbf{h}_k+\sigma_C^2}=\frac{\frac{\alpha_0^2P_T}{d_k^2K}}{\frac{\alpha_0^2P_T(K-1)}{d_k^2K}+\sigma_C^2}.
\end{equation}

If $K< N_t$, the linear precoder is redundant; hence, $K$ antennas can be directly manipulated to transmit $K$ ISAC signals individually. The received SINR is same as (\ref{omnisinr}).

In conclusion, the sum rate achievable by the IoT device $k$ during the $n$-th time slot is denoted as
\begin{equation}\label{sumrate}
R_{k,n}=\log_{2}{(1+\gamma_{k,n})}.
\end{equation}
\subsubsection{Radar Process}
According to \cite{7071108}, the echos of MIMO orthogonal waveform are still orthogonal, so radar can recognize the reflection of each device. Consequently, for device $k$, the received echo reflection is represented as
\begin{equation}
\mathbf{r}_{k,n}(t)=\beta_{k}\mathbf{b}(\theta_{k})\mathbf{a}^H(\theta_{k})\mathbf{x}_n(t-\tau_{k,n})+\mathbf{z}_{k,n}(t),
\end{equation}
where $\mathbf{b}(\theta_{k})$ is receive steering vector and has the same structure as $\mathbf{a}(\theta_{k})$, denoted as:
\begin{equation}
\mathbf{b}(\theta_k)=[1,e^{j2{\pi}\frac{d}{\lambda}\sin{\theta_k}},...,e^{j2{\pi}\frac{d}{\lambda}(N_r-1)\sin{\theta_k}} ]^T,
\end{equation}
and $\mathbf{z}_{k,n}(t)$ represents the zero-mean complex additive white Gaussian noise with variance $\sigma^2$. Both $\beta_{k,n}$ and $\tau_{k,n}$ signify the reflection coefficient and time lag for device $k$ during time slot $n$. The coefficient of reflection is expressed as
\begin{equation}
\beta_{k}=\frac{\varepsilon}{d_{k}^2},
\end{equation}
where $\varepsilon$ denotes the complex radar cross-section (RCS).

Using a conventional matched-filtering methodology \cite{6324717}, signal delays can be estimated and the refined output vectors are

\begin{equation}
\Tilde{\mathbf{R}}_{k,n}=G_m\beta_{k}\frac{P_T}{N_t}\mathbf{b}(\theta_{k})\mathbf{a}^H(\theta_{k})+\mathbf{Z}_r,
\end{equation}
where $G_m$ is signal processing gain, brought by matched-filtering. The noise matrix, $\mathbf{Z}_r$, consists of independent, zero-mean, and complex Gaussian elements with variance $\sigma_r^2$. Additionally, the model measuring the distance  is
\begin{equation}\label{estitau}
\hat{\tau}_{k,n}=\frac{2d_{k}}{c}+z_{\tau_{k,n}}.
\end{equation}
With delay, the distance $\hat{d}_{k,n}$ can be simply derived. Employing either the Capon method or the generalized likelihood ratio test (GLRT) \cite{4350230}, the angle between the ABS and device $k$ is deduced as
\begin{equation}\label{estitheta}
\hat{\theta}_{k,n}=\theta_{k,n}+z_{\theta_{k,n}},
\end{equation}
where $\theta_{k,n}$ is equivalent to $\theta_k$. The measurement models of (\ref{estitau}) and (\ref{estitheta}) both take the Gaussian distribution model. 

Accordingly, $z_{\tau_{k,n}}$ and $z_{\theta_{k,n}}$ both represent the Gaussian noise in the measurement with zero mean and variances $\sigma^2_{\tau_{k,n}}$ and $\sigma^2_{\theta_{k,n}}$, respectively. Given the challenge in acquiring variances for $\hat{\tau}_{k,n}$ and $\hat{\theta}_{k,n}$, the Cramér-Rao Bound (CRB) is utilized as the sensing metric, due to its unbiased estimation and ability to provide a lower MSE bound \cite{9705498}:
\begin{align}
\sigma^2_{\tau_{k,n}}&=CRB_\tau=\frac{\alpha_1}{SNR_{k,n}N_t N_r\kappa^2},\\
\sigma^2_{\theta_{k,n}}&=CRB_\theta=\frac{\alpha_2}{SNR_{k,n}N_t N_r\xi^2},
\end{align}
where pre-set constants are $\alpha_1$ and $\alpha_2$, related to system configurations. The radar echo signal-noise ratio (SNR) is denoted as $SNR_{k,n}=
\frac{P_T G_m |\beta_{k}|^2}{\sigma^2}$. As per \cite{9705498}, the squared effective bandwidth is $\kappa^2$ and the root mean square aperture width of the beampattern is
\begin{equation}
\xi^2=\frac{\pi^2 d^2_k \cos^2{\theta_k}(N_t^2-1)}{3\lambda^2}.
\end{equation}

With (\ref{estitau}) and (\ref{estitheta}), we can derive estimated channel $\hat{\textbf{h}}_{k,n}$, which is part of the optimization problem in Section III.
\subsection{Directional Phase}
Within this phase, the beampattern is tailored based on acquired parameters, enabling the transmission of a directional beampattern and the setting is established as $n=2,...,N$.
\subsubsection{Communication Process}
The received signal for device $k$ is represented as
\begin{equation}
y_{k,n}(t)=\mathbf{h}_{k,n}^H\mathbf{W}_n\mathbf{s}_n(t)+z_C(t).
\end{equation}
In concise terms, the SINR for device $k$ can be articulated as
\begin{align}\label{dirsinr}
\gamma_{k,n}=\frac{|\mathbf{h}_{k}^H \mathbf{w}_{k,n}|^2}{\sum_{i=1,i\neq k}^K |\mathbf{h}_{k}^H \mathbf{w}_{i,n}|^2+\sigma_C^2},
\end{align}
where the beamforming vector, denoted as $\mathbf{w}_{k,n}$, stands as the $k$-th column vector extracted from $\mathbf{W}_n$. For time slot $n$, the achievable sum rate pertaining to IoT device $k$ aligns with expression $(\ref{sumrate})$.
\subsubsection{Radar Process}
According to the assumption of \cite{9171304}, the steering vectors from different angles are asymptotically orthogonal under the massive MIMO regime, so the reflected echoes do not interfere with each other. Similarly, the received echo reflection is characterized as
\begin{equation}
\mathbf{r}_{k,n}(t)=\beta_{k}\mathbf{b}(\theta_{k})\mathbf{a}^H(\theta_{k})\mathbf{W}_n\mathbf{s}(t-\tau_{k,n})+\mathbf{z}_{k,n}(t).
\end{equation}
Post the implementation of matched-filtering, the models used for measurements remain consistent with $(\ref{estitau})$ and $(\ref{estitheta})$. Concludingly, the SNR of the radar's echoed signal can be delineated as 
\begin{align}
SNR_{k,n}=\frac{G_m |\beta_{k}|^2 \sum_{i=1}^K|\mathbf{a}^H(\theta_{k})\mathbf{w}_{i,n}|^2 }{\sigma^2}.
\end{align}

\section{Problem Formulation and Solution}\label{sec.train}
In this section, we develop a formulation for the beamforming problem employing sensing-assisted communication to optimize the communication rate for all devices.
\subsection{Problem Formulation}
The channel estimated by an ABS to device $k$ using position information from the radar echo is given by
\begin{equation}
\hat{\mathbf{h}}_{k,n}=\frac{\alpha_0}{\hat{d}_{k,n}}\mathbf{a}(\hat{\theta}_{k,n}).
\end{equation}
The ISAC system postulates that the communication rate with device $k$ at time slot $n$ is
\begin{align}
\hat{R}_{k,n}=\log_{2}{\left(1+\frac{\text{tr}(\hat{\mathbf{h}}_{k,n} \hat{\mathbf{h}}_{k,n}^H \mathbf{W}_{k,n})}{\sum_{i=1,i\neq k}^K \text{tr}(\hat{\mathbf{h}}_{k,n} \hat{\mathbf{h}}_{k,n}^H \mathbf{W}_{i,n})+\sigma_C^2)}\right)},
\end{align}
where the matrix $\mathbf{W}_{k,n}$ is denoted as $\mathbf{w}_{k,n}\mathbf{w}_{k,n}^H$. Therefore, the optimization problem can be expressed as
\begin{subequations}\label{opti1}
\begin{align}
&~~~~~~~~~~~~~~~~\underset{\{\mathbf{W}_{k,n}\}}{\text{max}}\sum\limits_{k=1}^K\hat{R}_{k,n} \label{opti1-a} \\
&~~~~~~~~~~~~~~~~\text{s.t.}~ \text{tr}\left(\sum\limits_{k=1}^K \mathbf{W}_{k,n}\right)=P_T \label{opti1-b} \\
&~~~~~~~~~~~~~~~~\mathbf{W}_{k,n}\succeq 0,\mathbf{W}_{k,n}=\mathbf{W}_{k,n}^H,~\forall k,\label{opti1-c} \\
&~~~~~~~~~~~~~~~~\text{rank}(\mathbf{W}_{k,n})=1,~\forall k, \label{opti1-d} \\
&~~~~~~~~~~~~~~~~\hat{R}_{k,n}\geq \Gamma_k,~\forall k,  \label{opti1-e}\\
&|\mathbf{a}^H(\hat{\theta}_{k,n-1})\mathbf{W}_{k,n}\mathbf{a}(\hat{\theta}_{k,n-1})-\mathbf{a}^H(\theta_{k,n}^{cover})\mathbf{W}_{k,n}\mathbf{a}(\theta_{k,n}^{cover})|  \nonumber \\
&~~~\leq B_k \text{tr} (\mathbf{W}_{k,n}),~\forall k,~\forall |\hat{\theta}_{k,n-1}-\theta_{k,n}^{cover}|\leq l\sigma_{\theta_{k,n-1}} \label{opti1-f},
\end{align}
\end{subequations}
where the transmission power limitation is imposed by equation (\ref{opti1-b}). The matrices $\mathbf{W}_{k,n}$ are subject to three constraints, which are semidefinite, Hermitian, and rank-one, as specified by equations (\ref{opti1-c}) and (\ref{opti1-d}). The term $\Gamma_k$ denotes the communication rate threshold for the $k$-th device, and increasing the value of  $\Gamma_k$ allows more power to be given to remote devices to ensure the minimum communication rate. Drawing from the findings in \cite{4350230}, we integrate sensing and communication paradigms within equation (\ref{opti1-f}), capturing the influence of sensing accuracy on beamforming strategies. Pertinently, within the context of equation (\ref{opti1-f}), based on the Gaussian distribution principle, the constant $l$ is generally chosen as 3 to cover the true angle as much as possible. This ensures maximum angle coverage corresponding to $\hat{\theta}_{k,n-1}$. Consequently, the cumulative angles $\theta_{k,n}^{cover}$, fulfilling the condition $|\hat{\theta}_{k,n-1}-\theta_{k,n}^{cover}|\leq l\sigma_{\theta_{k,n-1}}$, delineate the beam coverage. To achieve a compact transmission beam, the parameter $B_k$ is minimized. In short, by constraining the beam gain of the device at possible angles, (24f) makes the optimized beampattern an approximate ideal radar beampattern with width controllable.
\subsection{Problem Solution}
By ignoring rank-one constraint (\ref{opti1-d}), problem (\ref{opti1}) is rewritten as

\begin{align}
&\underset{\{\mathbf{W}_{k,n}\}}{\text{max}}\sum\limits_{k=1}^K\log_{2}{\left(1+\frac{\text{tr}(\hat{\mathbf{h}}_{k,n} \hat{\mathbf{h}}_{k,n}^H \mathbf{W}_{k,n})}{\sum_{i=1,i\neq k}^K \text{tr}(\hat{\mathbf{h}}_{k,n} \hat{\mathbf{h}}_{k,n}^H \mathbf{W}_{i,n})+\sigma_C^2)}\right)} \nonumber \\
\label{opti2}
&~~~~~~~~~~~~~\text{s.t.}~(\text{\ref{opti1-b}}),(\text{\ref{opti1-c}}),(\text{\ref{opti1-e}})~ \text{and} ~ (\text{\ref{opti1-f}}).
\end{align}

Secondly, we adopt SCA to approximate the non concave objective function in problem (\ref{opti2}) to a concave objective function, which is implemented through multiple iterations. We consider that iteration $q \geq 0$, and $\{ \mathbf{W}_{k,n} \}$ is depicted as $\{ \mathbf{W}_{k,n}^{(q)} \}$ accordingly. Afterwards, we have
\begin{align}
\label{sca1}
\hat{R}_{k,n}&=\log_{2}{\left( {\textstyle \sum_{i=1}^K} \text{tr}(\hat{\mathbf{h}}_{k,n} \hat{\mathbf{h}}_{k,n}^H \mathbf{W}_{i,n})+\sigma_C^2 \right)}\nonumber \\
&~~~~-\log_{2}{\left( {\textstyle \sum_{i=1,i\neq k}^K} \text{tr}(\hat{\mathbf{h}}_{k,n} \hat{\mathbf{h}}_{k,n}^H \mathbf{W}_{i,n})+\sigma_C^2 \right)} \\
&\ge \log_{2}{\left( {\textstyle \sum_{i=1}^K} \text{tr}(\hat{\mathbf{h}}_{k,n} \hat{\mathbf{h}}_{k,n}^H \mathbf{W}_{i,n})+\sigma_C^2 \right)}\nonumber \\
\label{sca2}
&~~~~-\log_{2}{\left( {\textstyle \sum_{i=1,i\neq k}^K} \text{tr}(\hat{\mathbf{h}}_{k,n} \hat{\mathbf{h}}_{k,n}^H \mathbf{W}_{i,n}^{(q)})+\sigma_C^2 \right)} \nonumber \\
&~~~~-{\textstyle \sum_{i=1,i\neq k}^K} \text{tr} \left(\mathbf{X}_{k,n}^{(q)}(\mathbf{W}_{i,n}-\mathbf{W}_{i,n}^{(q)})\right) \triangleq \tilde{R}_{k,n}^{(q)},
\end{align}
where $\mathbf{X}_{k,n}^{(q)}$ is defined as
\begin{align}
\mathbf{X}_{k,n}^{(q)}=\frac{\log_{2}{(e)} \hat{\mathbf{h}}_{k,n}\hat{\mathbf{h}}_{k,n}^H}{{\textstyle \sum_{i=1,i\neq k}^K} \text{tr}(\hat{\mathbf{h}}_{k,n} \hat{\mathbf{h}}_{k,n}^H \mathbf{W}_{i,n}^{(q)})+\sigma_C^2}.
\end{align}

In short, we implement first-order Taylor expansion on the second term of  (\ref{sca1}), then take its lower bound, and the obtained objective function in (\ref{sca2}) will approach convex function. Accordingly, problem (\ref{opti2}) is transformed into the following problem (\ref{opti2}) in the $q$-th iteration.
\begin{align}\label{opti3}
\underset{\{\mathbf{W}_{k,n}\}}{\text{max}}&\sum\limits_{k=1}^K\tilde{R}_{k,n}^{(q)} \nonumber \nonumber \\
\text{s.t.}~&(\text{\ref{opti1-b}}),(\text{\ref{opti1-c}}),(\text{\ref{opti1-e}})~ \text{and} ~ (\text{\ref{opti1-f}}).
\end{align}

Now the problem (\ref{opti3}) is convex, which can be optimally solved by convex optimization solvers such as CVX. After multiple iterations, assuming we obtain an approximate iterative solution of . If $ \{\mathbf{W}_{k,n}^{(q^\star)}\}$ satisfies the rank-one constraint, then it is the solution to the original problem (\ref{opti1}). However, in most cases, $ \{\mathbf{W}_{k,n}^{(q^\star)}\}$ is not a rank-one matrix, but a suboptimal solution can be obtained by classic SDR technique. 
Due to our strict requirements on the beam shape, conventional eigenvalue decomposition and Gaussian randomization \cite{8288677} cannot meet our needs, so we have decided to adopt IRM.
\section{Simulation Results}
In this section, we discuss the numerical findings to substantiate the efficacy of the introduced ISAC protocol. With the main reference to \cite{9171304} and \cite{9747255}, our simulation parameters delineated in Table \ref{parameter}. For the sake of consistency, we fixed the antenna spacing to half the wavelength, ensured that the number of transmitting antennas mirrors the number of receiving ones, and maintained an equal noise variance for both communication and radar, represented as $d = \lambda/2$, $N_t=N_r$, and $ \sigma_C^2 = \sigma^2$. 

\begin{table}[h]
\centering
\renewcommand{\arraystretch}{1.4}
\caption{Simulation Parameters}

\begin{tabular}{c|c}
 \hline 
\textbf{parameter} & \textbf{value}  \\
\hline\hline
Number of transmit antennas $N_t$ & 20   \\
\hline
Effective bandwidth $\kappa$ (Mhz)& 4    \\
\hline
Signal processing gain $G_m$ & 10   \\
\hline
Carrier wavelength $\lambda$ (m) & 0.06   \\
\hline
Additive white Gaussian noise $\sigma_C^2$ & 1   \\
\hline
Interval length $\Delta T$ & 0.01  \\
\hline
Power loss of reference distance $\alpha_0^2$ (dB) & 50  \\
\hline
\end{tabular}\label{parameter}
\end{table}

\begin{figure}[h] 
\centerline{\includegraphics[width=0.45\textwidth]{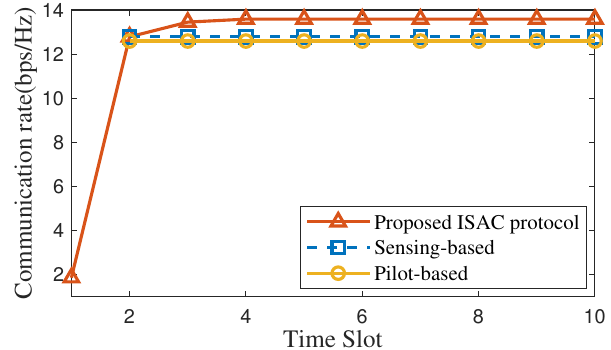}}
\caption{Proposed ISAC protocol versus the traditional omnidirectional sensing-based method and the pilot-based approach.}
\label{fig.vs}
\end{figure}

\begin{figure}[h] 
\centerline{\includegraphics[width=0.45\textwidth]{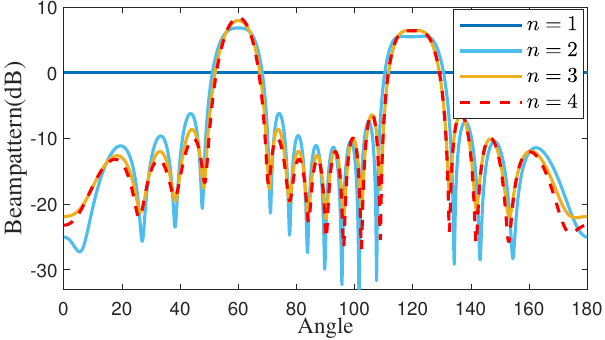}}
\caption{Beampattern at different time slots.}
\label{fig.timeslot}
\end{figure}
\begin{figure}[h] 
\centerline{\includegraphics[width=0.45\textwidth]{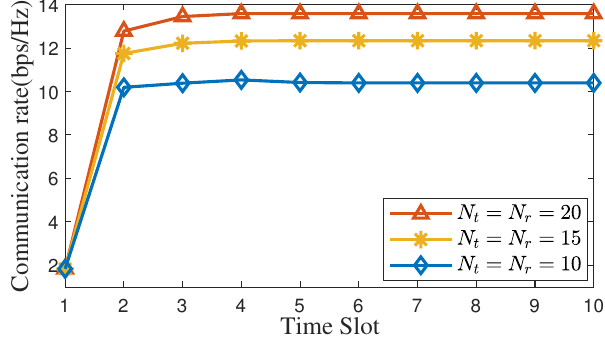}}
\caption{Communication rate performance of different number of antennas.}
\label{fig.antenna}

\end{figure}

Fig. \ref{fig.vs} provides an initial assessment, contrasting the communication rate attained through various techniques across time slots. Two devices are considered: Device 1 located at an angle and distance of 60$^\circ$ and 50m, respectively, and Device 2 positioned at 120$^\circ$ and 100m. For reference, 'Sensing-based' utilizes omnidirectional sensing in the initial phase and then transmits information through communication waveform. For the 'Pilot-based' method, linear minimum mean squared Error (LMMSE) channel estimation is used before communication. The communication rate for the proposed protocol is initially minimal but elevates as data accuracy improves. By the time slot $n$ = 4, the communication rate plateaus, evidencing an enhancement of 6.3$\%$ and 7.9$\%$ over 'Sensing-based' and 'Pilot-based', respectively. Over the duration of $N$ time slots, the throughput experiences an elevation of 5.7$\%$ and 8.7$\%$ compared to the aforementioned methods, respectively.

Fig. \ref{fig.timeslot} visually represents how the beampattern fluctuates across time slots. Retaining parameters from Fig. \ref{fig.vs}, at $n$ = 2, Device 1 exhibits a reduced beam coverage relative to Device 2, an outcome of its proximal location yielding heightened sensing precision. As time slots progress, the beam coverage visibly contracts, with the beampattern gain consistently amplifying. By the point of $n$ = 4, the beampattern becomes invariant.

The influence of varying antenna quantities on the beampattern is depicted in Fig. \ref{fig.antenna}. While keeping other parameters static, a reduction in antenna numbers correlates with a notable communication rate decline for every time slot. Relative to our proposed protocol, the communication rates for '$N_t$ = $N_r$ = 15' and '$N_t$ = $N_r$ = 10' decline by 9.8$\%$ and 29.2$\%$, respectively. Such reductions are attributed as follows. Primarily, the antenna quantity influences the variance in angle and distance measurements, which translates to measurement accuracy. Furthermore, a decrement in gain directly compromises the SINR of the radar echo, further influencing accuracy in subsequent time slots.
\section{Conclusion}

In this paper, we have proposed a sensing-assisted communication protocol based on the ISAC system.  The central proposition is the enhancement of the downlink communication rate by dynamically refining the transmit beampattern of the current time slot, capitalizing on sensory data procured from the preceding slot. Additionally, the integration of the SCA technique and the IRM algorithm has facilitated the resolution of the proposed optimization challenge. The numerical results validate the advantages of the proposed method. Further, the lower complexity algorithms is worth studying in the future.

\vspace{-0.1 cm}

\end{document}